%% file: paper.tex
\documentclass[runningheads]{llncs}
\usepackage{amsmath}
\usepackage{graphicx}
\usepackage{censor}
\usepackage{cite}
\usepackage{color}
\usepackage{hyperref}
\usepackage{outlines}
\usepackage{amsfonts}
\usepackage{pifont}
\usepackage{wasysym}
\usepackage{amssymb}
\usepackage{todonotes}
\usepackage{soul}
\usepackage{wrapfig}
\usepackage{booktabs}
\usepackage{float}
\usepackage{array}
\usepackage{makecell}
\usepackage{algorithmicx}
\usepackage{algorithm}
\usepackage[noend]{algpseudocode}
\usepackage{paralist} 
\usepackage[T1]{fontenc}
\usepackage{tikz}
\usetikzlibrary{arrows, calc, shapes,petri}
\usepackage{extarrows}
\usepackage{comment}
\usepackage{wrapfig}

\newtheorem{assumption}{Assumption}

\input{sections/macros}

\begin{document}

\title{Detecting Dynamic Relationships in Object-Centric Event Logs}
\titlerunning{Detecting Dynamic Relationships in Object-Centric Event Logs}

\author{
 Alessandro Gianola\inst{1} \and
Zeeshan Hameed\inst{2} \and
 Marco Montali\inst{2} \and
Anjo Seidel\inst{3} \and
 Mathias Weske\inst{3} \and
 Sarah Winkler\inst{2} 
}

\authorrunning{A.~Gianola et al.}

\institute{
    INESC-ID/Instituto Superior Técnico, Universidade de Lisboa, Portugal \\
    \email{alessandro.gianola@tecnico.ulisboa.pt} \and  Free University of Bozen-Bolzano, Italy\\
    \email{\{hameed,winkler,montali\}@inf.unibz.it} \and
    Hasso Plattner Institute, University of Potsdam, Germany\\
    \email{\{anjo.seidel,mathias.weske\}@hpi.de}
}
\sloppy 
\maketitle

\input{sections/0_abstract}

\input{sections/1_introduction}

\input{sections/3_OCEL_semantics}
\input{sections/4_evaluation}
\input{sections/5_discussion}


\bibliographystyle{splncs04}
\bibliography{paper.bib}


\end{document}

%% file: sections/macros.tex
\newcommand{\objects}{\mathcal O}
\newcommand{\objtypes}{\Sigma}

\newcommand{\activities}{\mathcal A}

\newcommand{\m}[1]{\mathsf{#1}}

\tikzstyle{class}=[draw, rectangle, inner sep=1.5pt, line width=.7pt, scale=1, minimum width=24mm, minimum height=8mm]

\tikzstyle{place}=[draw, circle, inner sep=1.5pt, line width=.7pt, scale=.8, minimum width=6mm]
\tikzstyle{trans}=[draw, rectangle, inner sep=1.5pt, line width=.7pt, scale=.8, minimum width=6mm, minimum height=6mm, fill=gray!10]
\tikzstyle{silentTrans}=[draw, rectangle, inner sep=1.5pt, line width=.7pt, scale=.8, minimum width=2.5mm, minimum height=6mm, fill=black]
\tikzstyle{arc}=[draw, ->, line width=.5pt]
\tikzstyle{fatarc}=[draw, ->, line width=.5pt, double]
\tikzstyle{fatline}=[draw, -, line width=.5pt, double]
\tikzstyle{action}=[scale=.6]
\tikzstyle{starttoken}=[regular polygon, regular polygon sides=3,minimum width=3mm,fill=black,inner sep=0pt,rotate=30]
\tikzstyle{endtoken}=[regular polygon, regular polygon sides=4,minimum width=3mm,fill=black,inner sep=0pt]
\tikzstyle{pseudostart}=[regular polygon, regular polygon sides=3,minimum width=3mm,draw,inner sep=0pt, rotate=30]
\tikzstyle{pseudoend}=[regular polygon, regular polygon sides=4,minimum width=3mm,inner sep=0pt, draw]

\colorlet{frameColor}{yellow!60!white!40}
\colorlet{wheelColor}{red!60!white!50}
\colorlet{handleColor}{blue!70!white!30}
\colorlet{wheelFrameColor}{orange!50}
\colorlet{handleFrameColor}{green!100!white!30}
\tikzstyle{framePlace}=[place,fill=frameColor]
\tikzstyle{wheelPlace}=[place,fill=wheelColor]
\tikzstyle{handlePlace}=[place,fill=handleColor]
\tikzstyle{wheelFramePlace}=[place,fill=wheelFrameColor]
\tikzstyle{handleFramePlace}=[place,fill=handleFrameColor]
\colorlet{ocolor}{blue!70!green!20}
\colorlet{icolor}{yellow!90!red!20}
\colorlet{oicolor}{green!90!blue!20}
\tikzstyle{oplace}=[place,fill=ocolor]
\tikzstyle{otrans}=[trans,fill=ocolor]
\tikzstyle{iplace}=[place,fill=icolor]
\tikzstyle{itrans}=[trans,fill=icolor]
\tikzstyle{oiplace}=[place,fill=oicolor]
\tikzstyle{idplace}=[place,fill=red!30] 
\tikzstyle{dplace}=[place,fill=magenta!20]             
\tikzstyle{oidplace}=[place,fill=black!20]
\tikzstyle{oitrans}=[trans,fill=oicolor]
\tikzstyle{mixtrans}=[trans,shading = axis,rectangle, left color=ocolor, right color=icolor,shading angle=135]
\tikzstyle{oimixtrans}=[trans,shading = axis,rectangle, left color=ocolor, right color=oicolor,shading angle=135]
\tikzstyle{insc}=[scale=.8]
\tikzstyle{splitme}=[rectangle split, rectangle split horizontal,rectangle split parts=2]
\tikzstyle{log}=[fill=white]
\tikzstyle{model}=[fill=gray!10]

\newcommand{\eid}{\#}
\newcommand{\rf}{\mathit{ref}}

\newcommand{\manytype}{\mu}
\newcommand{\onetype}{\sigma}

%% file: sections/0_abstract.tex
\begin{abstract}
Object-centric process mining examines how processes interact with multiple co-evolving objects, and has gained great interest in recent years. However, object-centric event logs (OCELs) leave object relationships underspecified in several respects, especially if relationships are dynamic, i.e., they  change over time.
In this paper, we identify and formally define for the first time assumptions that allow to represent and manipulate dynamic relationships in OCELs in a semantically unambiguous way.
We evaluate existing logs to show that our assumptions are often satisfied, ensuring full transparency of relationship semantics. 
\keywords{Business Process Management  \and Dynamically Changing Relations \and Synchronization}
\end{abstract}

%% file: sections/1_introduction.tex
\section{Introduction}
\label{sec:intro}



Process mining aims to extract actionable insights from the execution data recorded by information systems~\cite{vanderaalstProcessMiningManifesto2012a}. By analyzing event logs, it seeks to reconstruct the actual behavior of operational processes and reason about performance, deviations, and conformance. While traditional process mining assumes that events refer to a single case notion, modern information systems store data about multiple interacting process instances that share co-evolving objects~\cite{vanderaalstObjectCentricProcessMining2019}. This has motivated the development of object-centric process mining, which studies processes that simultaneously
evolve
several interrelated objects of different types. Executions are recorded using the standard OCEL format (object-centric event logs~\cite{ghahfarokhiOCELStandardObjectCentric2021}), where each event may reference multiple objects.

 A variety of modeling formalisms has been proposed to represent object-centric processes: object-centric Petri nets (OCPN)~\cite{vanderaalstDiscoveringObjectcentricPetri2020} extend classical Petri nets to handle multiple object types. OCPNs are relatively simple and amenable to discovery, but known to be underspecified as they cannot track object identities and  relationships~\cite{Aal23}.
More expressive formalisms capturing some form of synchronization include synchronous proclets~\cite{fahlandDescribingBehaviorProcesses2019},
object-centric process trees~\cite{vanDetten2025},
and object-centric Petri nets with identifiers \cite{gianolaObjectCentricConformanceAlignments2024}.
While these formalisms can enforce synchronization semantics, they almost exclusively focus on \emph{stable} relationships between objects that remain unmodified over time.

The current literature thus almost entirely leaves out the important case where relationships dynamically change over time, despite the fact that such relationships are ubiquitous in real life processes: for example, a package might be moved from one truck to another, the destination address of a shipment can be dynamically updated, or the content of an order can change. 


This shortcoming in current research is largely due to the fact that OCEL, as well as the standard of object-centric event data (OCED)~\cite{oced}, neither relate object relationships with events, nor with time. In this paper, we illustrate the resulting ambiguity 
by a number of examples.
We then identify and formally define for the first time assumptions regarding how to log in an OCEL object-centric event data with dynamically changing relationships in an unambiguous way, with minimal additional domain knowledge. By evaluating available OCELs, we show 
that these assumptions are indeed respected to a great extent.


%% file: sections/3_OCEL_semantics.tex
\section{OCELs and Dynamic Relationships}
\label{sec:semantics}

Contemporary object-centric event data representations~\cite{ghahfarokhiOCELStandardObjectCentric2021} lack an explicitly defined semantics regarding the way events impact the evolution of object-to-object relationships
 --- that is, which are the temporal intervals in which links between objects 
hold. For clarity, we use \emph{relationship} at the type level, and \emph{link} at the instance level. E.g., the type-level information that ``an order contains items'' constitutes a relationship, while the fact that ``order $o_1$ contains item $i_2$'' indicates a link. 
Such relationships are in the literature almost always assumed to be \emph{rigid} (or \emph{stable}), that is, once established, a link never ceases to exist \cite{bertiOCELObjectCentricEvent2024,Multiple_Behavioral_Dimensions_EKG,ER2025}. 
%

In this paper we call a \emph{dynamic many-to-one} relationship between  a many-type $\manytype$ and a one-type $\onetype$ a relationship in which at every point in time, every object of type $\manytype$ is linked to at most one object of type $\onetype$. 
In contrast, in \emph{many-to-many} relationships between types $\manytype$ and $\onetype$, an object of type $\manytype$ can at some point be linked to many objects of type $\onetype$ and vice versa. 
In the sequel, \emph{many-to-one} relationships are always supposed to be dynamic. 

The goal of this section is to provide a minimally intrusive mechanism to unambiguously view how an OCEL tracks the evolution of links of relationships. Here ``minimally intrusive'' means that we want to limit, as much as possible, the number of assumptions made, and the additional domain knowledge needed.

For completeness, we recall the notion of object-centric event logs 
\cite{vanderaalstDiscoveringObjectcentricPetri2020}:

\begin{definition}
Given sets of object types $\objtypes$, activities $\activities$, and timestamps $\mathbb{T}$, an \emph{object-centric event log} (OCEL) is a tuple $L = ( E, \objects, \pi_{act}, \pi_{obj}, \pi_{time} )$ where:
\begin{inparaenum}[\itshape (i)]
    \item $E$ is a set of events;
    \item $\objects$ is a set of object identifiers typed by the function $ot\colon \objects \to \objtypes$;
    \item the functions $\pi_{act}\colon E \to \activities$, $\pi_{obj}\colon E \to \mathcal{P}(\objects)$, and $\pi_{time}\colon E \to \mathbb{T}$, mapping each event to an activity, a set of objects, and a timestamp.
\end{inparaenum}
\end{definition}

First and foremost, there are two possible ways of interpreting events in an OCEL, as illustrated by the following example. 

\begin{example}
\label{exa:dynamics}
Consider the situation where an order $o_1$ is created with items $i_1$ and $i_2$, then $i_2$ gets removed from $o_1$, and later $i_3$ is added. Depending on how object participation in events is interpreted, this can be expressed in two ways. 

In what we call the \emph{parameter semantics}, objects denote the actual parameters of the activity referred by the event. 
An OCEL could thus look as follows:\\[.5ex]
\begin{minipage}{0.39\textwidth}
\begin{tabular}{|l|l|ll|}
\hline
event & activity & order & items\\ \hline
$\eid_1$ & $\mathsf{create\ order}$ & $o_1$ & $i_1, i_2$ \\ \hline
$\eid_2$ & $\mathsf{reduce\ order}$ & $o_1$ & $i_2$  \\ \hline
$\eid_3$ & $\mathsf{add\ to\ order}$ & $o_1$ & $i_3$  \\ \hline
\end{tabular}
\end{minipage}
\begin{minipage}{0.59\textwidth}
Note that each event
is associated with the objects to which the respective activity applies. The current state after each event, in particular regarding which links are established resp. removed, remains however implicit.
\end{minipage}\\[.5ex]
On the other hand, one can adopt a so-called  \emph{snapshot semantics}, where the objects listed in each event do not represent parameters, but provide instead a snapshot of the new state  (e.g., the new state of an order) \emph{after} the respective activity, making the existing links explicit. This would yield the following log:\\[.5ex]
\begin{minipage}{0.39\textwidth}
\begin{tabular}{|l|l|ll|}
\hline
event & activity & order & items\\ \hline
$\eid_1$ & $\mathsf{create\ order}$ & $o_1$ & $i_1, i_2$ \\ \hline
$\eid_2$ & $\mathsf{reduce\ order}$ & $o_1$ & $i_1$  \\ \hline
$\eid_3$ & $\mathsf{add\ to\ order}$ & $o_1$ & $i_1, i_3$  \\ \hline
\end{tabular}
\end{minipage}
\begin{minipage}{0.58\textwidth}
Note how in event $\eid_{2}$, the associated item is $i_1$, i.e., the \emph{remaining} item in $o_1$ -- witnessing that the set of active links for the order-item relationship consists only of $(o_1,i_1)$. This is in contrast to the table above,  where event $\eid_{2}$ is
\end{minipage}\\[.5ex]
 associated with $i_2$, the item \emph{to which the removal operation is applied}.
\end{example}

When adopting parameter semantics, 
it is impossible to reconstruct from an OCEL alone which links hold at a given point in time: additional transactional domain knowledge on the updates semantics of activities is needed \cite{ChCM25}. Indeed, even though this is left implicit, existing approaches typically adopt the snapshot semantics \cite{Multiple_Behavioral_Dimensions_EKG}. 
We hence follow this assumption.

\begin{assumption}
\label{ass:snapshot}
    Events are recorded under the snapshot semantics, that is, listing which objects are related as an effect of the event.
\end{assumption}

However, the next example shows that even under the snapshot semantics, there is inherent ambiguity in the way OCELs capture 
changing
relationships. 

\begin{example}
\label{exa:one-to-many}
    Consider a process where employees handle orders, and events recording two order creations followed by a change of the handling employee:\\[.5ex]
\begin{minipage}{0.55\textwidth}
\begin{tabular}{|l|l|cc|}
\hline
 event & activity & order & employee \\ \hline
$\eid_4$ & $\mathsf{create\ order}$ & $o_1$ & $p_1$ \\ \hline
$\eid_5$ & $\mathsf{create\ order}$ & $o_2$ & $p_1$ \\ \hline
$\eid_6$ & $\mathsf{change\ order\ manager}$ & $o_1$  & $p_2$  \\ \hline
\end{tabular}
\end{minipage}
\begin{minipage}{0.44\textwidth}
From the co-participation of orders and employees to the same event, we can only infer that there exists a relationship between some of their  instances, but we do not know with
\end{minipage}\\[.5ex]
 which multiplicity constraints. 
This extra-knowledge is essential: If the relationship is many-to-many, then we may infer from the OCEL that order $o_1$ is linked to both employees $p_1$ and $p_2$ after $\eid_6$. 
If instead the relationship is many-to-one -- more specifically, every employee can handle multiple orders, but every order is handled by one and only one employee -- then we must instead infer that $o_1$ was initially linked to $p_1$, and upon  $\eid_6$ it changed its link to $p_2$. 
This witnesses that the relationship is a \emph{dynamic} one-to-many relationship instead of a rigid many-to-many relationship: over time the same order can be linked to distinct employees but in every instant to only one. 
\end{example}

Ex.~\ref{exa:one-to-many} shows that some minimal domain knowledge is needed regarding the multiplicity of relationships. In addition, many-to-many relationships are known to make it difficult to properly handle synchronization of related objects, as noted in~\cite{artaleModelingReasoningDeclarative2019,fahlandDescribingBehaviorProcesses2019}. As pointed out in \cite{fahlandDescribingBehaviorProcesses2019}, synchronization can be handled naturally by reification of every many-to-many relationship into a pair of many-to-one relationships with a newly introduced object type in between.
We thus assume that such reification was done in a preprocessing step.
Moreover, since a one-to-one relationship can be represented as two many-to-one relationships, we restrict ourselves in the sequel to many-to-one relationships, which directly support synchronization.
We also assume that relationships exist only between distinct types, since OCEL cannot distinguish objects of the same type with different roles.

\begin{assumption}
\label{ass:multiplicity}
We assume that domain knowledge is provided, indicating which pairs of distinct object types are connected by a many-to-one relationship.
\end{assumption}
In a many-to-one relationship, we refer to the ``one'' side as the parent, and to the ``many'' side as the child,
and when referring to a many-to-one relationship from the ``one'' side, we call it a one-to-many relationship.
We do not enforce mandatory participation on either side.
%
Even when assuming snapshot semantics, Ex.~\ref{exa:dynamics} and~\ref{exa:one-to-many} show that the reconstruction of links after an event may depend on whether the event primarily operates on the parent or on the child endpoint of a one-to-many relationship. In Ex.~\ref{exa:dynamics}, events primarily operate on the \emph{order} type, i.e., the parent side of the one-to-many relationship with \emph{item}. To unambiguously trace how the relationship evolves after an event that operates on both orders and items, it needs to list the set of items to which the order is linked in that instant. For example, when $o_1$ occurs with $i_1$ in $\eid_{2}$,
this implies that the item set $i_1, i_2$ linked to $o_1$ in $\eid_{1}$ is
replaced: link $(o_1,i_1)$ is kept, while $(o_1,i_2)$ removed.

Also in Ex.~\ref{exa:one-to-many}, events primarily operate on the \emph{order} type. However, there \emph{order} corresponds to the child side of the many-to-one relationship with \emph{employee}. Differently from the previous case, now every event that operates on both an order and an employee with primary focus on the order, indicates to which (parent) employee that order gets linked. Hence, \emph{other} orders that the same employee handles are out of scope, and not mentioned. For example, the fact that employee $p_1$ handles a second order $o_2$ according to event $\eid_{4}$ does \emph{not} mean that
$p_1$ does no longer handle order $o_1$ 
i.e., $\eid_{4}$ does not delete the link
between $p_1$ and $o_1$ established by $\eid_{1}$.

This shows that the snapshot semantics must be further clarified: in the first case, the active links for an order to its items can be reconstructed 
from the current event alone,
while in the second case the active links for an employee to its orders require to also consider previous events. These two ways of operating over links correspond to the classical alternative to maintain links from the many or the one side of the relationship. One may be tempted to impose that only one of the two conventions should be adopted in an OCEL. However, this cannot be done when events operate 
over three or more object types, as illustrated next.

\begin{example}
\label{exa:reference}
Consider an OCEL obtained by merging 
those
in Ex.~\ref{exa:dynamics} and~\ref{exa:one-to-many}:\\
\begin{minipage}{0.58\textwidth}
\begin{tabular}{|l|l|ccc|}
\hline
 & activity & order & item & employee \\ \hline
$\eid_1$ & $\mathsf{create\ order}$ & $o_1$ & $i_1, i_2$ & $p_1$ \\ \hline
$\eid_2$ & $\mathsf{reduce\ order}$ & $o_1$ & $i_1$ & \\ \hline
$\eid_3$ & $\mathsf{add\ to\ order}$ & $o_1$ & $i_1, i_3$ &  \\ \hline
$\eid_4$ & $\mathsf{create\ order}$ & $o_2$ & $i_4, i_5$ & $p_1$ \\ \hline
$\eid_5$ & $\mathsf{wrap\ item}$ & & $i_4$ &  \\ \hline
$\eid_6$ & $\mathsf{change\ order\ manager}$ & $o_1$ &  & $p_2$  \\ \hline
$\eid_7$ & $\mathsf{ship\ order}$ & $o_1$ & $i_1, i_3$ &  \\ \hline
$\eid_8$ & $\mathsf{ship\ order}$ & $o_2$ & $i_4, i_5$ &  \\ \hline
\end{tabular}
\end{minipage}
\ 
\begin{minipage}{0.40\textwidth}
Now orders are at once linked to the contained items and the employees that handle them. To properly interpret which active links exist 
after
an event, we need to know the primary object type (which we call \emph{reference type}) for each activity. 
In the OCEL here, this is \emph{order} for
\end{minipage}\\[.5ex]
all events except $\mathsf{wrap\ item}$,
which can only have reference type \emph{item}.
Given these reference types and the multiplicity of all relationships, one can reconstruct the active links after each event. Notice that for events of type $\mathsf{create\ order}$, we need to consider that the order-to-item relationship is treated from the parent side (thus enumerating the set of items to which the mentioned order is linked), while the order-to-employee one is treated from the child side (thus only indicating what is the current employee to which the mentioned order is linked). 


\end{example}

\noindent
Generalizing the observations in Ex.~\ref{exa:reference}, we get to the following assumptions.

\begin{assumption}
\label{ass:reference}
\label{cond:single_ref_object}
Every activity has one object type as its \emph{reference type},
and every event has exactly one object of the reference type of its activity. We call this object the \emph{reference object}.
\end{assumption}

\begin{assumption}
\label{ass:links}
 Let $e=(a,O_e)$ be an event in the log with activity $a$ and object set $O_e$, and $t_\rf$ the reference type of $a$.
For every object $o \in O_e$ that has an object type $t$  that is by the domain knowledge in a relationship with $t_\rf$,
either
\begin{compactenum}[(a)]
\item
\label{cond:many_to_one}
$(t_\rf,t)$ is a many-to-one relationship, $o$ is the only object in $O_e$ of type $t$, and $o$ is the only object currently in relationship with the reference object; or
\item
\label{cond:one_to_many}
$(t_\rf,t)$ is a one-to-many relationship and the subset of $O_e$ with type $t$ are all objects 
of type $t$ currently in relationship with the reference object.
\end{compactenum}
\end{assumption}

These conventions clarify the confusion in Ex.~\ref{exa:reference}:
if for all activities except $\mathsf{wrap\ item}$, 
we choose the reference type \emph{order}, each event mentions only one object of its reference type, so Assumption \ref{cond:single_ref_object} holds.
As the relationship \emph{order -- item} is one-to-many, when activities with reference type \emph{order} refer to sets of items, these are \emph{all} items currently in the order (Assumption~\ref{ass:links} \ref{cond:one_to_many}).
On the other hand, as \emph{order -- employee} is many-to-one, if activity $\m{create\ order}$ with reference type \emph{order} mentions employee $p_1$, this is the parent of the relationship, and an event mentioning another order $o_2$ with $p_1$ does not delete the link between $o_1$ and $p_1$; however, when in event $\eid_6$ the order $o_1$ occurs with another employee $p_2$, we assume that the link $(o_1, p_1)$ gets deleted (Assumption~\ref{ass:links}\ref{cond:many_to_one}).
%



A last point of attention pertains to a ``locality'' principle for activities: 
when an event occurs, what is the scope of the changes its corresponding activity can induce? Intuitively, one may expect that this scope contains only links of the reference object.
However, to ensure that children in a one-to-many relationship have only one parent at a time, (implicit) parent updates may be required, which may break this locality principle, as witnessed by the following example.

\begin{example}
\label{exa:implicit:deletion}
Consider the following OCEL recording teams of multiple employees, with a many-to-one relationship $\mathit{employee} - \mathit{team}$, so that an employee can be a member of only one team (reference types are underlined):\\
\begin{minipage}{0.45\textwidth}
\begin{tabular}{|l|l|cc|}
\hline
 & activity & team & empl. \\ \hline
$\eid_1$ & $\mathsf{create\ \underline{team}}$
  & $t_1$ & $p_1, p_2$  \\ \hline
$\eid_2$ & $\mathsf{create\ \underline{team}}$
  & $t_2$ & $p_3, p_4$ \\ \hline
$\eid_3$ & $\mathsf{add\ \underline{employee}\ to\ team}$
  & $t_2$ & $p_2$ \\ \hline
$\eid_4$ & $\mathsf{create\ \underline{team}}$
  & $t_3$ & $p_1, p_3$ \\ \hline
\end{tabular}
\end{minipage}
\begin{minipage}{0.5\textwidth}
According to Assumptions~\ref{ass:reference} and \ref{ass:links},  after $\eid_3$ all of $p_2, p_3, p_4$ belong to team $t_2$ (note that the reference type of $\eid_3$ is \emph{employee}).
However, $p_2$ is at this point already a member of team $t_1$.
\end{minipage}
Therefore, to maintain the multiplicity constraints of the many-to-one relationship, we need to conclude that the relationship $(t_1, p_2)$ is implicitly deleted by $\eid_3$.
A similar problem occurs in $\eid_4$, where employees $p_1$ and $p_3$ are added to a team, even if they are already a member of another one: also here an implicit deletion would be required to maintain the one-to-many relationship.
However, while the implicit deletion in $\eid_3$ pertains to the former parent of the reference object, the implicit deletions needed in $\eid_4$ pertain to (former) parents of children of the reference object. So the latter case violates the locality principle.
\end{example}
Ex.~\ref{exa:implicit:deletion} shows that if the locality principle is desired, it must be enforced as an additional assumption
to prohibit implicit deletions as in $\eid_4$ of Ex.~\ref{exa:implicit:deletion}:

\begin{assumption}
\label{ass:implicit:deletion}
Each event only modifies links that involve its reference object.
\end{assumption}

%% file: sections/4_evaluation.tex
\section{Evaluation}
\label{sec:evaluation}

To evaluate the practicality of our  interpretation of OCELs, we check the five assumptions presented in Sec. \ref{sec:semantics} as much as possible against publicly available event logs\footnote{The evaluation script in \url{https://github.com/bytekid/dynamic-relationships} was applied to OCELs from \url{https://www.ocel-standard.org/event-logs}.}. 
The evaluation of \autoref{ass:snapshot} on snapshot semantics requires domain knowledge, and is hence not possible from event logs alone. However, according to the textual descriptions of the considered logs, no information contradicting the snapshot semantics was found. 
\autoref{ass:multiplicity} requires cardinalities of relationships to be provided. Again, for a reliable interpretation domain knowledge is required. However, assuming the log to be sufficiently complete, similar as required by Armstrong relations in database theory~\cite{Armstrong}, by counting co-occurrences of objects in events, an educated guess can be made. The respective relationship counts for one-to-one (oo), many-to-one (mo), and many-to-many (mm) are listed in \autoref{tab:logs}, for both static relationships and dynamic relationship candidates. For static relationships, co-occurrences were counted across the entire log~\cite{ER2025}; for the dynamic case, we only counted objects co-occurring in the same events.
The differences between these two counts show that dynamic relations are in fact quite common.
Next, for \autoref{ass:reference} we verify whether every event type has a candidate for a reference type, i.e., an object type of which in every event of this type exactly one object occurs. In \autoref{tab:logs}, entries of the form $x/y$ indicate that for $x$ out of $y$ event types reference types could be identified, and the respective percentage.
We also give the number of possible combinations of reference types (rt). 
In absence of domain knowledge  on the update semantics of events, \autoref{ass:links} provides an interpretation of changing relationships but cannot be seen as violated or respected from the log alone.
Finally, for \autoref{ass:implicit:deletion}, for OCELs where possible reference types were identified earlier, we consider the detected dynamic many-to-one and one-to-one relationships to check whether events only modify links that involve their reference objects. The ratios of events satisfying this assumption are listed in the last column of \autoref{tab:logs}, for the  combination of reference types where this value is highest (for AoE not all combinations could be checked due to their vast number, though).

\begin{table}[t]
    \caption{The assumptions evaluated on publicly available OCELs.}
    \centering
\begin{tabular}{l|c|cc|ccc|c}
OCEL & $\#\objtypes$ 
& static  & dynamic 
& \multicolumn{3}{c|}{Assumpt. 3} 
& \multicolumn{1}{c}{Assumpt. 5}\\
& &  oo/mo/mm &  oo/mo/mm & & & rt & \\
\hline
Logistics & 7 & 0/3/6 & 6/1/3 & 14/14 &(100\%) & 846 & 100\%\\
Order Man. & 6 & 0/5/8 & 3/7/3 & 11/11 &(100\%) & 1296 & 82\%\\
P2P & 7 & 2/6/1 & 3/5/1 & 10/10 &(100\%) & 128 & 92\%\\
Angular Github & 2 & 0/0/1 & 0/0/1 & 16/67 &(24\%) & -- & --\\
Hinge Production & 12 & 3/19/10 & 29/3/0 & 11/11 &(100\%) & 746496 & 99\%\\
AoE & 30 & 0/69/66 & 62/60/13 & 851/851 &(100\%) & $> 10^{370}$ & $\geq$68\%\\
LRMS-O2C & 9 & 1/9/1 & 10/1/0 & 20/22 &(91\%) & -- & --\\
LRMS-P2P & 6 & 0/0/8 & 7/1/0 & 12/12 &(100\%) & 432 & 100\%\\
LRMS-Hiring & 6 & 0/4/4 & 7/1/0 & 24/24 &(100\%) & 3072 & 100\%\\
LRMS-Hospital & 8 & 0/4/5 & 9/0/0 & 9/9 &(100\%) & 288 & 100\%\\
Bundestag & 44 & 0/0/31 & 7/16/8 & 95/148 &(64\%) & -- & --\\
Inventory Man. & 7 & 0/7/10 & 17/0/0 & 30/30 &(100\%) & $\approx10^{18}$ & 100\%\\
\end{tabular}
\label{tab:logs}
\end{table}

%% file: sections/5_discussion.tex
\section{Related Work and Conclusion}
\label{sec:discussion}

\noindent 
\textit{Related work.}
So far, virtually all mining/analysis approaches dealing with OCELs support explicit relationships between objects \cite{bertiOCELObjectCentricEvent2024}, or implicit relationships established when they co-participate to the same event \cite{Multiple_Behavioral_Dimensions_EKG}.
However, such relationships are almost always assumed to be stable.
In~\cite{lissTOTeMTemporalObject2024}, temporal object type models (TOTeM) are mined from OCELs, but the dynamics in these models consider temporal existence windows of objects while their relationships are considered stable: a relationship is simply supposed to exist in all time instants where both objects exist. 
In \cite{vanDetten2025} instead, \emph{temporal implications} are proposed as one form of dynamic relationship.
This notion captures that for a certain timespan, a fixed set of objects of a certain type is linked to only one set of objects of another type. 
This  implicitly assumes snapshot semantics (\autoref{ass:snapshot}), as the linked objects are supposed to be those that occurred together in an event.
Moreover, temporal implications 
are a rather restricted form of dynamic relationships, e.g. orders with a growing set of products cannot be modeled, as sets of objects are only compared wrt. equality. 
Temporal implications basically require stable relationships for a fixed block-structured sub-process;

\noindent
\textit{Conclusion.\ }
We identified key assumptions to uniquely reconstruct the evolution of dynamic relationships from an input OCEL, given additional minimal domain knowledge on the cardinalities of object type relationships. 
The evaluation of OCELs showed that these assumptions are indeed often satisfied.
In future work, these assumptions can be valuable to discover process models that explicitly capture and enforce the evolution of dynamic relationships, as well as to shape the representation of object relationships in future versions of  OCED~\cite{oced}.

%% file: paper.bib
@string{lncs = "LNCS"}

@string{lnbip = "LNBIP"}

@string{springer = "Springer"}

@inproceedings{Aal23,
  author       = {{van der Aalst}, Wil M. P.},
  title        = {Toward More Realistic Simulation Models Using Object-Centric Process
                  Mining},
  booktitle    = {Proc. 37th {ECMS}},
  pages        = {5--13},
  nopublisher    = {ECMS},
  year         = {2023},
}

@article{oced,
  author       = {Dirk Fahland and
                  Marco Montali and
                  Julian Lebherz and
                  Wil M. P. van der Aalst and
                  Maarten van Asseldonk and
                  Peter Blank and
                  Lien Bosmans and
                  Marcus Brenscheidt and
                  Claudio Di Ciccio and
                  Andrea Delgado and
                  Daniel Calegari and
                  Jari Peeperkorn and
                  Eric Verbeek and
                  Lotte Vugs and
                  Moe Thandar Wynn},
  title        = {Towards a Simple and Extensible Standard for Object-Centric Event
                  Data {(OCED)} - Core Model, Design Space, and Lessons Learned},
  journal      = {CoRR},
  volume       = {abs/2410.14495},
  year         = {2024},
  nodoi          = {10.48550/ARXIV.2410.14495},
  eprinttype    = {arXiv},
  eprint       = {2410.14495},
  bibsource    = {dblp computer science bibliography, https://dblp.org}
}

@misc{bertiOCELObjectCentricEvent2024,
  title = {{{OCEL}} ({{Object-Centric Event Log}}) 2.0 {{Specification}}},
  author = {Berti, Alessandro and et al.},
  year = {2024},
  number = {arXiv:2403.01975},
  eprint = {2403.01975},
  primaryclass = {cs},
  publisher = {arXiv},
  urldate = {2024-10-02},
  archiveprefix = {arXiv}
}

@inproceedings{fahlandDescribingBehaviorProcesses2019,
  title = {Describing {{Behavior}} of {{Processes}} with {{Many-to-Many Interactions}}},
  booktitle = "Proc. 40th PETRINETS",
  _booktitle = {Application and {{Theory}} of {{Petri Nets}} and {{Concurrency}}},
  author = {Fahland, Dirk},
  noeditor = {Donatelli, Susanna and Haar, Stefan},
  year = {2019},
  pages = {3--24},
  series = lncs,
  volume = 11522,
  publisher = springer,
  abstract = {Processes are a key application area for formal models of concurrency. The core concepts of Petri nets have been adopted in research and industrial practice to describe and analyze the behavior of processes where each instance is executed in isolation. Unaddressed challenges arise when instances of processes may interact with each other in a one-to-many or many-to-many fashion. So far, behavioral models for describing such behavior either also include an explicit data model of the processes to describe many-to-many interactions, or cannot provide precise operational semantics.},
  isbn = {978-3-030-21571-2},
  langid = {english},
  keywords = {Many-to-many interactions,Modeling,Multi-instance processes,Petri nets,True-concurrency semantics},
  file = {/Users/anjo.seidel/Zotero/storage/UNRZW3VH/Fahland - 2019 - Describing Behavior of Processes with Many-to-Many Interactions.pdf}
}

@inproceedings{ghahfarokhiOCELStandardObjectCentric2021,
  title = {{{OCEL}}: {{A Standard}} for {{Object-Centric Event Logs}}},
  shorttitle = {{{OCEL}}},
  booktitle = {Proc. ADBIS 2021},
  author = {Ghahfarokhi, Anahita Farhang and et al.},
  year = {2021},
  pages = "169--175",
  publisher = springer,
  isbn = {978-3-030-85082-1},
  langid = {english}
}

@inproceedings{gianolaObjectCentricConformanceAlignments2024,
  title = {Object-{{Centric Conformance Alignments}} with~{{Synchronization}}},
  booktitle = {Proc. 36th CAiSE},
  author = {Gianola, Alessandro and Montali, Marco and Winkler, Sarah},
  date = {2024},
  year = {2024},
  nopublisher = springer,
  series = lncs,
  volume = 14663,
  pages = "3--19",
  isbn = {978-3-031-61057-8},
  langid = {english}
}

@inproceedings{lissTOTeMTemporalObject2024,
  title = {{{TOTeM}}: {{Temporal Object Type Model}} for~{{Object-Centric Process Mining}}},
  shorttitle = {{{TOTeM}}},
  booktitle = {Proc. BPM Forum},
  author = {Liss, Lukas and Adams, Jan Niklas and {van der Aalst}, Wil M. P.},
  year = {2024},
  publisher = springer,
pages = "107--123",
  isbn = {978-3-031-70418-5},
  langid = {english}
}

@inproceedings{ChCM25,
  author       = {Rikayan Chaki and
                  Diego Calvanese and
                  Marco Montali},
  noeditor       = {Khalid Saeed and
                  Jir{\'{\i}} Dvorsk{\'{y}} and
                  Makoto Fukumoto and
                  Nobuyuki Nishiuchi},
  title        = {Generation of Timelines from Event Knowledge Graphs Using Domain Knowledge},
  booktitle    = {Proc. 24th {CISIM}},
  series       = lncs,
  volume       = {15927},
  pages        = {275--289},
  nopublisher    = {Springer},
  year         = {2025},
  nourl          = {https://doi.org/10.1007/978-3-032-02406-0\_20},
  nodoi          = {10.1007/978-3-032-02406-0\_20},
  bibsource    = {dblp computer science bibliography, https://dblp.org}
}

@book{Multiple_Behavioral_Dimensions_EKG,
  author    = {Dirk Fahland},
  title     = {Process Mining over Multiple Behavioral Dimensions with Event Knowledge Graphs},
  booktitle = {Process Mining Handbook},
  editor    = {Wil M.P. van der Aalst and Josep Carmona},
  series    = lnbip,
  volume    = {448},
  pages     = {231--255},
  nopublisher = {Springer},
  year      = {2022},
  nodoi       = {10.1007/978-3-031-08848-3_9},
  address   = {Cham}
}

@article{vanderaalstDiscoveringObjectcentricPetri2020,
  title = {Discovering {{Object-centric Petri Nets}}},
  author = {{van der Aalst}, Wil M. P. and Berti, Alessandro},
  year = {2020},
  journal = {Fundamenta Informaticae},
  volume = {175},
  number = {1-4},
  publisher = {IOS Press},
  issn = {0169-2968},
  urldate = {2024-10-02},
  langid = {english},
  pages = "1--40"
}

@inproceedings{vanderaalstObjectCentricProcessMining2019,
  title = {Object-{{Centric Process Mining}}: {{Dealing}} with {{Divergence}} and {{Convergence}} in {{Event Data}}},
  shorttitle = {Object-{{Centric Process Mining}}},
  booktitle = {Software {{Engineering}} and {{Formal Methods}}},
  author = {{van der Aalst}, Wil M. P.},
  year = {2019},
  publisher = springer,
  pages = "3--25",
  isbn = {978-3-030-30446-1},
  langid = {english}
}

@inproceedings{vanderaalstProcessMiningManifesto2012a,
  title = {Process {{Mining Manifesto}}},
  booktitle = {Proc. BPM Workshops},
  author = {{van der Aalst}, Wil M. P. and et al.},
  year = {2012},
  publisher = {Springer},
  noaddress = {Berlin, Heidelberg},
  isbn = {978-3-642-28108-2},
  langid = {english}
}

@inproceedings{vanDetten2025,
  title = {Modeling and Discovering Dynamic Identity
Relations in Object-Centric Process Mining},
  booktitle = {Proc. 7th ICPM},
  author = {{van Detten}, Jan Niklas and Schumacher, Pol and Leemans, Sander},
  year = {2025},
  pages = "",
}

@incollection{artaleModelingReasoningDeclarative2019,
  title = {Modeling and {{Reasoning}} over {{Declarative Data-Aware Processes}} with {{Object-Centric Behavioral Constraints}}},
  booktitle = {Proc. 17th BPM},
  author = {Artale, Alessandro and Kovtunova, Alisa and Montali, Marco and Van~Der Aalst, Wil M. P.},
  noeditor = {Hildebrandt, Thomas and Van Dongen, Boudewijn F. and R{\"o}glinger, Maximilian and Mendling, Jan},
  year = 2019,
  volume = {11675},
  pages = {139--156},
  publisher = springer,
  noaddress = {Cham},
  nodoi = {10.1007/978-3-030-26619-6_11},
  series       = lncs,
  urldate = {2024-01-30},
  isbn = {978-3-030-26618-9 978-3-030-26619-6},
  langid = {english}
}

@inproceedings{ER2025,
  author       = {Anjo Seidel and
                  Sarah Winkler and
                  Alessandro Gianola and
                  Marco Montali and
                  Mathias Weske},
  noeditor       = {Dominik Bork and
                  Roman Lukyanenko and
                  Shazia Sadiq and
                  Ladjel Bellatreche and
                  Oscar Pastor},
  title        = {To Bind or Not to Bind? {D}iscovering Stable Relationships in Object-Centric
                  Processes},
  booktitle    = {Proc. 44th {ER}},
  series       = lncs,
  volume       = {16189},
  pages        = {223--241},
  year         = {2025},
  nodoi          = {10.1007/978-3-032-08623-5\_12},
}

@inproceedings{Armstrong,
  title={Dependency Structures of Data Base Relationships},
  author={William Ward Armstrong},
  booktitle={Proc. IFIP Congress},
  year={1974},
  nourl={https://api.semanticscholar.org/CorpusID:38788061}
}
